\documentclass{article}
\usepackage{spconf,amsmath,graphicx}
\usepackage{footnote}
\usepackage{url}
\usepackage{booktabs}
\usepackage{hyperref}
\usepackage{amsmath}
\usepackage{etoolbox}
\usepackage{appendix}

\AtBeginEnvironment{equation}{\vspace{-2ex}}

\title{SynVox2: Towards a privacy-friendly VoxCeleb2 dataset}
\name{\shortstack{Xiaoxiao Miao$^1$, Xin Wang$^1$, Erica Cooper$^1$, Junichi Yamagishi$^1$,\\ 
\vspace{-0.1cm}
Nicholas Evans$^2$, Massimiliano Todisco$^2$, Jean-François Bonastre$^3$, Mickael Rouvier$^3$}}
\address{\shortstack{$^1$National Institute of Informatics, Japan \\
\vspace{-0.1cm}
 $^2$Digital Security Department, EURECOM, France 
 $^3$LIA, University of Avignon, France}}
\begin{document}
\ninept
\onecolumn
{\noindent\Large \textbf{IEEE Copyright Notice}}

${}$

{\noindent\large \copyright 2024 IEEE. 
Personal use of this material is permitted. Permission from IEEE must be obtained for all other uses, in any current or future media, including reprinting/republishing this material for advertising or promotional purposes, creating new collective works, for resale or redistribution to servers or lists, or reuse of any copyrighted component of this work in other works.

${}$

\noindent
This work has been submitted to the IEEE International Conference on Acoustics, Speech and Signal Processing.
}
\twocolumn
\maketitle
\begin{abstract}
The success of deep learning in speaker recognition relies heavily on the use of large datasets.
However, the data-hungry nature of deep learning methods has already being questioned on account the ethical, privacy, and legal concerns that arise when using large-scale datasets of natural speech collected from real human speakers. 
For example, the widely-used VoxCeleb2 dataset for speaker recognition is no longer accessible from the official website. 
To mitigate these concerns, this work presents an initiative to generate a privacy-friendly synthetic VoxCeleb2 dataset that ensures the quality of the generated speech in terms of privacy, utility, and fairness. We also discuss the challenges of using synthetic data for the downstream task of speaker verification.
\end{abstract}
\section{Introduction}
\label{sec:intro}
Large-scale speech data and powerful computing resources are key to the success of deep learning methods in automatic speaker verification (ASV).
However, the use of speech as a form of biometric data is governed by a set of legal restrictions, such as the General Data Protection Regulation (GDPR) \cite{GDPR}. 

Related legal and ethical issues have already led to the withdrawal of well-known large-scale datasets used for face recognition research, namely the  VGGFace2 \cite{cao2018vggface2} and MS-Celeb-1M \cite{guo2016ms} databases both of which were constructed by crawling facial images from the web.
Researchers have hence begun to explore the potential of using synthetic images for face recognition research \cite{qiu2021synface,boutros2022sface,boutros2023synthetic}.
The ASV field has faced similar problems due to privacy issues. For instance, the widely-used large-scale VoxCeleb2 dataset \cite{chung2018voxceleb2}, which contains speech data collected from  5,994 speakers, has become a standard ASV benchmark, though the database is \emph{no longer available} from the official website\footnote{\url{https://www.robots.ox.ac.uk/~vgg/data/voxceleb/vox2.html}}. The withdrawal of other popular biometric databases will likely soon follow. It is then inevitable that the community will have no choice but to consider alternatives.

\begin{figure}[t]
\centering
\includegraphics[width=0.8\columnwidth]{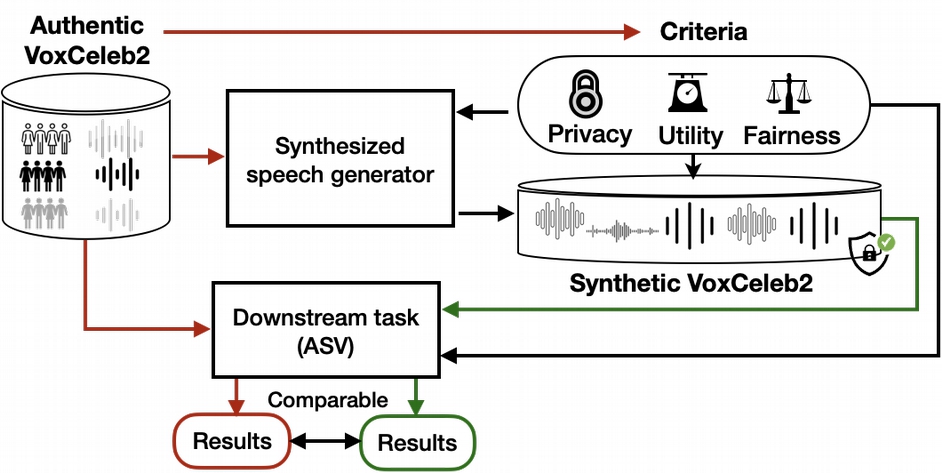}
 \vspace{-0.3cm}
\caption{Overview of the creation of the synthetic Voxceleb2 dataset. The scenario assumes a reliable party holds authentic data for generating and evaluating shareable synthetic data, with the assurance that the authentic data will not be released.}
\label{fig:overview}
 \vspace{-0.5cm}
\end{figure}

As synthetic data is a promising option, this work aims to explore the creation and utilization of a privacy-friendly synthetic VoxCeleb2 dataset for ASV model training.
Figure \ref{fig:overview} illustrates the general idea of the synthetic VoxCeleb2 called SynVox2. 
The original VoxCeleb2, which is also referred to as \emph{authentic VoxCeleb2}, is fed to a speech generator to create a \emph{synthetic VoxCeleb2} database, which can meet two primary criteria: adequately protect speaker privacy while maintaining utility comparable to the authentic database. One sub-criteria is to mitigate data bias and increase fairness.
The synthetic VoxCeleb2 dataset is subsequently used to train ASV models, with the goal of achieving comparable results to the same models trained with the authentic VoxCeleb2.

One possible solution to protect speaker privacy while still allowing the sharing of speech data is through speaker anonymization
\cite{tomashenko2021voiceprivacy, tomashenko2022voiceprivacy, miao22_odyssey,miao2022analyzing,miao2023language,meyer2023anonymizing}. 
This method hides the speaker's identity (privacy) while preserving other speech characteristics, such as content and emotion. 
Anonymization techniques could also be used to generate synthetic speaker voices and data. By using speaker anonymisation for this purpose, we aim to generate privacy-friendly synthetic dataset, which can strike a balance between protecting privacy and supporting research.

In this work, we employ the recently proposed language-robust orthogonal Householder neural network (OHNN)-based speaker anonymization technique \cite{miao2023language} to create a privacy-friendly VoxCeleb2 dataset called SynVox2. 
With the same number of speakers as the authentic VoxCeleb2 dataset, SynVox2 can be shared with far fewer privacy concerns compared to the sharing of the authentic VoxCeleb2 database. 
In addition, we define several metrics for evaluating the use of SynVox2 in terms of privacy, utility, and fairness. 
These metrics may serve as a protocol for future research, enabling researchers to assess whether a synthetic dataset is suitable for their ASV research. 
Furthermore, we conduct an in-depth analysis of intra-/inter-speaker variations in SynVox2, aiming to improve the utility of SynVox2.
Specifically, we propose methods for increasing intra-/inter-speaker variations, such as modeling background noise from authentic speech and incorporating it into the generated speech.

\section{Requirements for a privacy-friendly synthetic speech database}
\label{sec:require}
This section describes the three requirements a privacy-friendly synthetic speech database should satisfy: privacy, utility, and fairness. It also introduces the evaluation metrics used to assess the degree to which these requirements are fulfilled. 

\subsection{Ensuring Privacy through Unlinkability}
Privacy-sensitive information in speech extends beyond that related to speaker identity. Nonetheless, privacy can still be preserved to a great extent by obfuscating the speaker identity since any remaining privacy-sensitive content cannot be linked to the original speaker. We hence consider a privacy-friendly synthetic speech database to be able to protect the speaker's identity. The privacy of a speaker's identity in a synthetic database is protected if it is \emph{unlinkable} to its original identity in the authentic database \cite{NAUTSCH2019441}\footnote{Again, the authentic database will not be released in our assumed scenario, shown in Figure \ref{fig:overview}}. Given two speech samples of a speaker from the synthetic and the authentic databases, respectively, privacy is protected if it is difficult to determine that the two samples have been uttered by the same speaker.

This study evaluates privacy protection performance using an ASV evaluation model. The enrollment data is from the authentic VoxCeleb2 database, while the test trial is from SynVox2, as shown in Figure \ref{fig:flowchart}.
A privacy-friendly speech database should achieve a as high as possible ASV equal error rate (EER), 
which indicates that the ASV evaluation model has difficulty linking an authentic enrollment utterance and a protect test utterance.

\subsection{Maintaining Data Utility}
\begin{figure}[t]
\centering
\includegraphics[width=0.7\columnwidth]{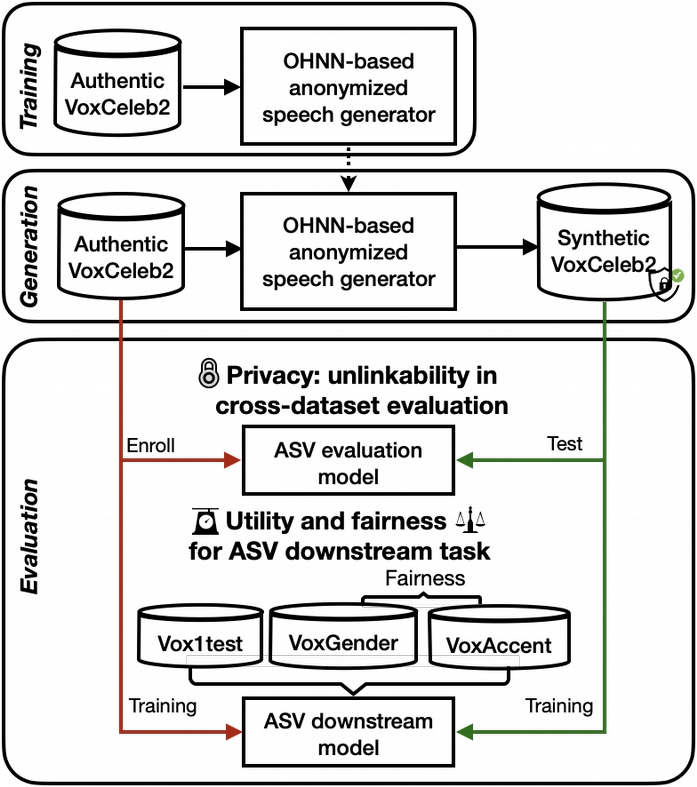}
 \vspace{-0.3cm}
\caption{Flowchart of OHNN-based synthetic VoxCeleb2 generation and evaluation.}
\label{fig:flowchart}
 \vspace{-0.4cm}
\end{figure}

Downstream models trained on a privacy-friendly synthetic database with high utility are expected to perform similarly to models trained using authentic data.
This study considers ASV as a downstream task, shown to the bottom part of Figure \ref{fig:flowchart}.
Utility is assessed in terms of the ASV EER which is estimated from experiments performed on authentic test sets.
ASV models trained using either authentic VoxCeleb2 or SynVox2 should exhibit similar performance.

\subsection{Reducing Data Bias and Increasing Fairness}
While data utility mainly measures the overall downstream performance on test sets, it is essential to ensure that downstream models trained on a privacy-friendly synthetic database do not disfavor any particular group in the test set, e.g., genders, dialects, and ethnicities. 

With ASV as the downstream task, this study uses the Fairness Disrepancy Rate (FDR) \cite{de2021fairness,estevez2023study} to assess the fairness. 
Given decision threshold $\tau$, the FDR considers the largest distance between false alarm rates $\text{FAR}$ (i.e., non-target speaker trials being classified as target) and false reject rates $\text{FRR}$ (i.e., target trials being classified as non-target) over multiple groups.
Given a set of groups $D = \{d_1,d_2,...,d_n\}$, 
the FDR is defined as:
\begin{equation} \label{eq1}
\setlength{\jot}{-5pt}
\begin{split}
\text{FDR} &= 1 - \Big(\alpha \times \max(\lvert \text{FAR}^{d_i}(\tau) - \text{FAR}^{d_j}(\tau) \rvert) \\
    &\quad + (1 - \alpha) \times \max(\lvert \text{FRR}^{d_i}(\tau) - \text{FRR}^{d_j}(\tau) \rvert)\Big).
\end{split}
\end{equation}\\[-3ex]
where $\text{FAR}^{d_i}$ and $\text{FAR}^{d_j}$ are the false alarm rates for groups $d_i$ and $d_j$, $\forall d_i, d_j \in D$, respectively. $\text{FRR}^{d_i}$ and $\text{FRR}^{d_j}$ are the false reject rates of groups $d_i$ and $d_j$, respectively. $\alpha \in [0, 1]$ is a design choice and set to $0.95$ so as to give greater importance to disparities in the rate of more costly false alarms than to the rate of false rejects.
$\text{FDR}=1$ means the system is perfectly fair. $\text{FAR}^{d_i}$, $\text{FAR}^{d_j}$, $\text{FRR}^{d_i}$, and $\text{FRR}^{d_j}$ of each subgroup are calculated using a fixed decision threshold $ \tau = 10e-1$, when considering trials of all the groups together.

We study fairness across gender and accent groups.
The gender group consists of female and male speakers.
The accent group includes English speakers haling from Austria, Canada, France, Germany, India, Italy, the Netherlands, Spain, the UK and the USA.

\begin{figure}[t]
\centering
\includegraphics[width=0.9\columnwidth]{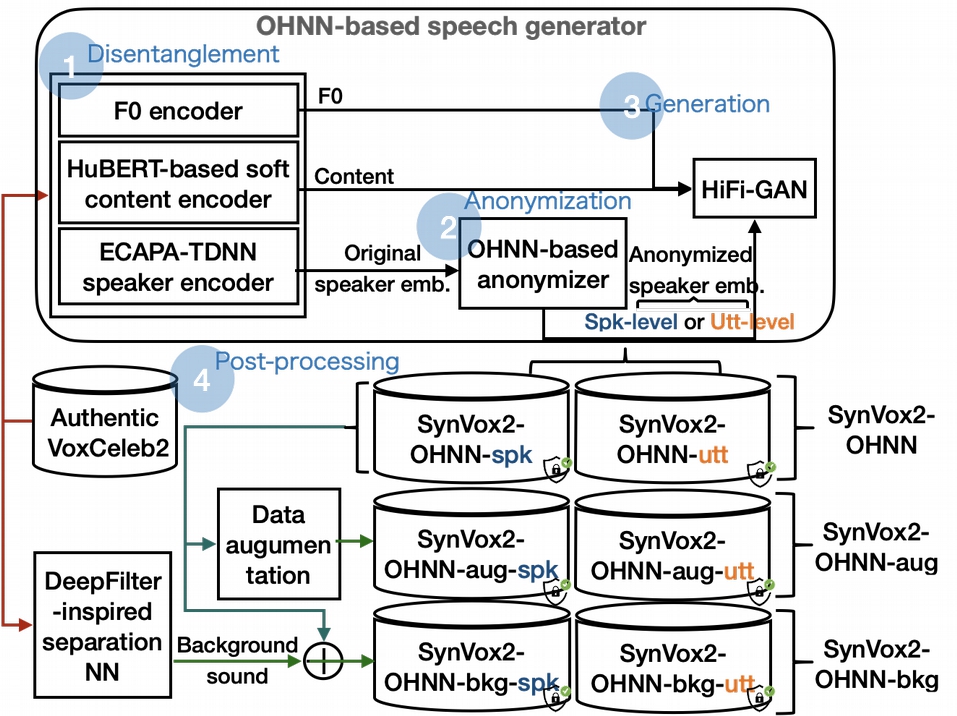}
 \vspace{-0.3cm}

\caption{Architecture of OHNN-based anonymized speech generator.}
\label{fig:method}
 \vspace{-0.5cm}
\end{figure}

\section{SynVox2 Generation Methods}
\label{sec:methods}
This section presents the process to create a privacy-friendly synthetic version of the VoxCeleb2 database that satisfies the constraints and criteria presented in section \ref{sec:require}.

\subsection{Language-robust OHNN-based speaker anonymization}
\label{sec:ohnn}
Speaker anonymization is one approach to create privacy-friendly synthetic datasets.
Because VoxCeleb2 is a multilingual dataset, we use a recently proposed language-robust OHNN-based anonymization method \cite{miao2023language} to generate different speaker-anonymised versions of VoxCeleb2. 
It uses a self-supervised learning (SSL)-based content encoder and an OHNN-based anonymizer, which supports the generation of speaker-distinctive anonymized speech even in languages unseen in training. The generation process involves three steps, as shown to the top of Figure \ref{fig:method}:

\textit{1) Disentanglement:}
The YAAPT algorithm \cite{kasi2002yet} is used to extract F0.
The ECAPA-TDNN speaker encoder is trained on the \textit{VoxCeleb2} \cite{chung2018voxceleb2} datasets and provides 192-dimensional speaker identity representations.
The HuBERT-based soft content encoder is fine-tuned on \textit{LibriTTS-train-clean-100} \cite{zen2019libritts} from a pre-trained HuBERT Base model\footnote{\url{https://github.com/pytorch/fairseq/tree/main/examples/hubert}} to capture the speech contents. 

\textit{2) Anonymization:}
The OHNN-based anonymizer \cite{miao2023language} rotates the original speaker embeddings to corresponding anonymized speaker embeddings using multiple orthogonal Householder transformation \cite{householder1958unitary} layers. 
The weights of the OHNN are randomly initialized and trained with classification and distance losses that prevent anonymized speakers from overlapping with other original and anonymized speakers. Hence, 
the anonymized speaker embeddings generated by the trained OHNN-based anonymizer are expected to have distinctive pseudo-speaker identities. 

\textit{3) Generation:}
Finally, the content features, F0, and anonymized speaker embeddings are passed to a HiFi-GAN model \cite{kong2020hifi} for audio waveform generation.
The HiFi-GAN model is trained using the \textit{LibriTTS-train-clean-100} database \cite{zen2019libritts}.

\vspace{-2mm}
\subsection{Synthetic Datasets}
\vspace{-1mm}

The dataset generated directly by the OHNN-based speech generator is referred to as \textit{SynVox2-OHNN}, as illustrated at the top of Figure \ref{fig:method}.

Because VoxCeleb2 was collected under diverse real-world conditions and contains various types of background noise, the OHNN-based speech generator pre-trained solely on clean data may not reproduce the background noise of the authentic VoxCeleb2 database. 
This would reduce the variations among utterances in \textit{SynVox2-OHNN}. 
To alleviate the reduced intra/inter-speaker variation, we generate different versions of SynVox2 using the post-processing methods shown at the bottom of Figure \ref{fig:method}.

\noindent
\textit{1) SynVox2-OHNN-aug}: 
One straightforward approach is directly adding noise, reverberation to \textit{SynVox2-OHNN}.
Similar to the standard speech data augmentation method \cite{ko2017study}, we use room impulse responses and background noise from \cite{ko2017study} and the MUSAN database \cite{musan2015}.
Although this seems to be redundant to the data augmentation in downstream ASV training, our experiments demonstrated that ASV downstream models benefited from the double augmentation.

\noindent 
\textit{2) SynVox2-OHNN-bkg}:
Inspired by techniques used for preserving background sound in voice conversion \cite{yao2023preserving}, we use a pre-trained DeepFilter-inspired model \cite{schroter2022deepfilternet} to separate background sound and clean speech from authentic speech.
The background sound is then added to \textit{SynVox2-OHNN} to ensure consistent ambient characteristics. 
DeepFilterNet-inspired model was developed in-house and trained on a large-scale dataset comprising noise extracted from YouTube videos and clean speech.

Another factor affecting the intra-speaker variations stems from the manner in which we extract speaker embeddings. 
One strategy involves extraction of embeddings by grouping together the set of all utterances corresponding to each speaker (denote \emph{-spk} in Figure \ref{fig:method}).
Although this strategy ensures a consistent pseudo speaker identity, it results in the reduced intra-speaker variations. We hence explore an alternative utterance-level approach (denote \emph{-utt} in Figure \ref{fig:method}), in which anonymized speaker embeddings are extracted individually from each authentic input utterance. The anonymized embeddings extracted from utterances corresponding to the same speaker may then differ and better capture intra-speaker variability.
We explored both approaches each combined with the two post-processing approaches.

\begin{table}[t!]
\caption{EERs (\%) achieved by the ASV evaluation model in cross-dataset evaluation.
EER in first row is calculated on authentic VoxCeleb2 dataset for reference.
Other rows are results in various cross-dataset conditions.
Higher EERs in cross-dataset conditions indicate that the synthetic and authentic speakers are hardly associated. }
\label{tab:unlink}
\centering
\begin{tabular}{llll}
\toprule
Cross datasets                 &  EER(\%) \\ \midrule
Authentic vs Authentic         & \phantom{0}2.07   \\ \midrule  
Authentic vs. SynVox2-OHNN-spk      &  32.66   \\ 
Authentic vs. SynVox2-OHNN-aug-spk  &   34.43  \\  
Authentic vs. SynVox2-OHNN-bkg-spk  &   27.84    \\ \midrule  
Authentic vs. SynVox2-OHNN-utt      & 32.81     \\ 
Authentic vs. SynVox2-OHNN-aug-utt  &  34.76   \\ 
Authentic vs. SynVox2-OHNN-bkg-utt  & 27.81  \\ 
\bottomrule
\end{tabular}
\end{table}

\begin{figure}[t!]
\centering
 \vspace{-0.2cm}
\includegraphics[width=0.8\columnwidth]{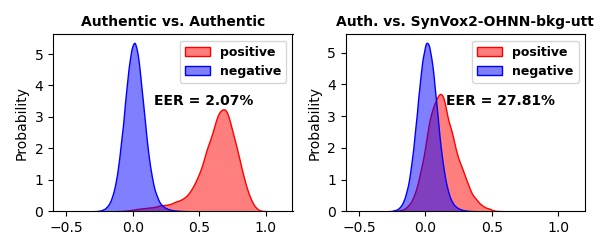}
 \vspace{-0.5cm}
\caption{EERs (\%) and score distribution achieved by ASV evaluation model in cross-dataset evaluation for authentic VoxCeleb2 and SynVox2-OHNN-bkg-utt datasets.}
\label{fig:unlink}
\vspace{-0.5cm}
\end{figure}

\begin{figure*}[t]
    \centering
    \includegraphics[width=1.4\columnwidth, height=0.4\columnwidth]{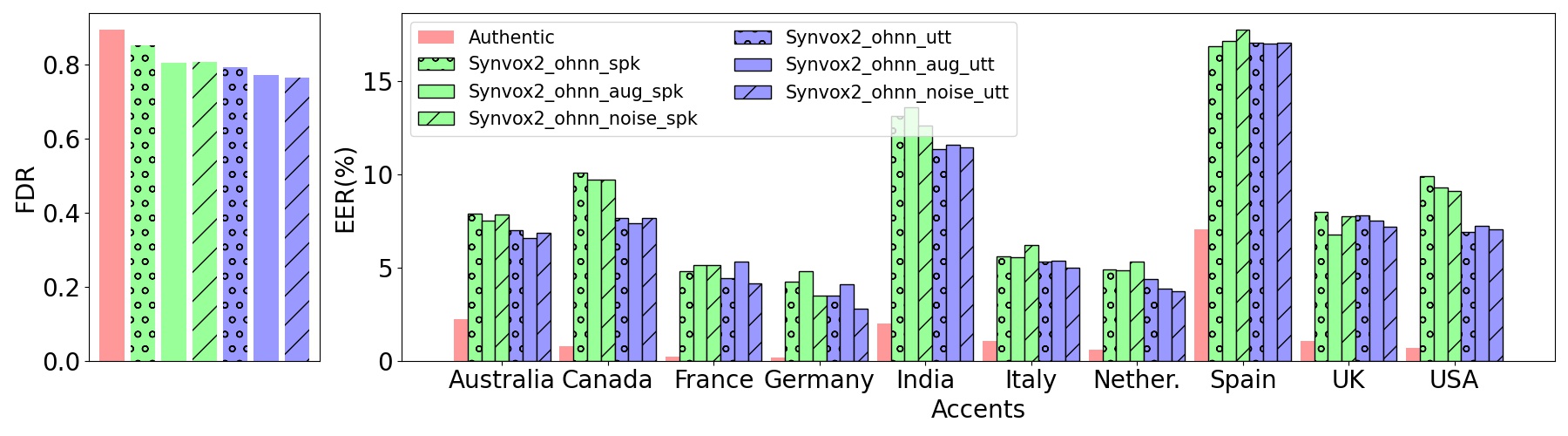}
    \hspace{0.2cm}
    \includegraphics[width=0.6\columnwidth, height=0.4\columnwidth]{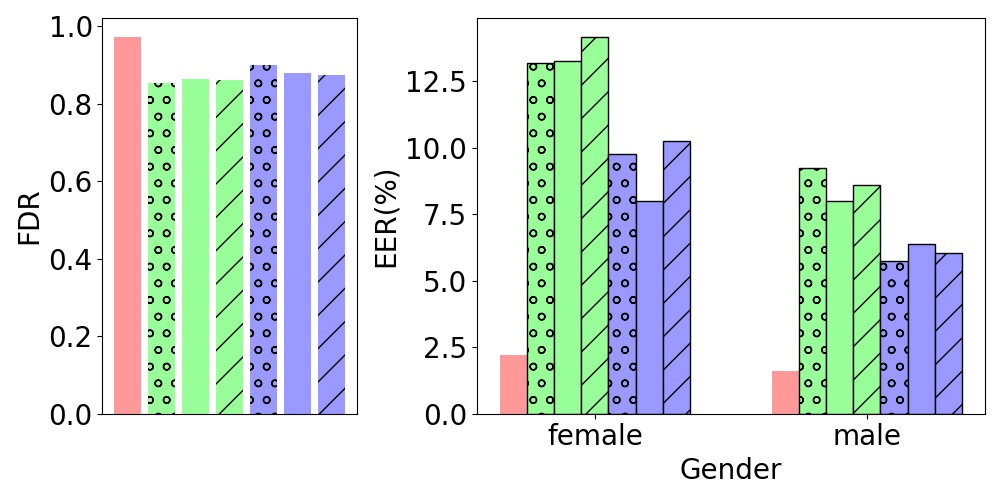}
     \vspace{-0.5cm}
    \caption{EER and FDR for accent (right two plots) and gender (left two plots) groups, respectively.}
    \label{fig:combined_fair}
     \vspace{-0.5cm}
\end{figure*}

\begin{figure}[t!]
\centering
\includegraphics[width=0.9\columnwidth]{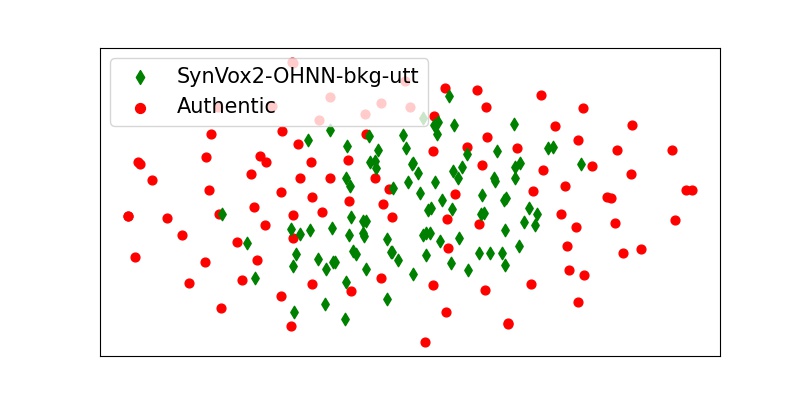}
 \vspace{-0.6cm}
\caption{ Visualization of speaker embeddings for samples from authentic VoxCeleb2 and SynVox2-OHNN-bkg-utt datasets. Each point corresponds to a distinct speaker.}
\label{fig:oo_aa}
 \vspace{-0.6cm}
\end{figure}

\section{Experiments}
\subsection{Setup}
\textit{OHNN Training}: The language-robust OHNN-based anonymized speech generator is trained on authentic VoxCeleb2 used random orthogonal Householder reflections, with a random seed of 50 for parameter initialization. 
We used an additive loss function which combines weighted angular margin softmax and cosine similarity. 
Full details of the training procedure can be found in \cite{miao2023language}.

\noindent
\textit{SynVox2 Evaluation}: 
\textbf{Unlinkability}: We generate enrollment-test pairs, where enrollment utterances are sourced from the authentic VoxCeleb2 database, whereas test utterances are selected from the SynVox2 database. Specifically, for each speaker, we randomly selected four utterances to form two same-speaker trials. For each speaker, we also generated 100 different-speaker trials by pairing one of the 4 utterances with 100 utterances of different speakers. This resulted in a total of $611388=(2+100) * 5994$ trials. We used a publicly available ECAPA-TDNN\footnote{\url{https://huggingface.co/speechbrain/spkrec-ecapa-voxceleb}} ASV system designed using the SpeechBrain toolkit \cite{speechbrain} to compute the EER.
\textbf{Utility}:
First, to establish a benchmark, we trained a downstream ASV model (ECAPA-TDNN with 512 channels in the convolution frame layers \cite{desplanques2020ecapa}) using the SpeechBrain recipe with data augmentation.
Then, we trained multiple alternative downstream ASV models using the same recipe as the benchmark ASV model, but with each of the SynVox2 databases
The utility of each model was then estimated in terms of the EER using the test partition of the VoxCeleb1 database~\cite{nagrani2017voxceleb}.
\textbf{Fairness}:
We built gender and accent test sets from the VoxAccent dataset \cite{estevez2023study} to further assess fairness. VoxAccent is a subset of the VoxCeleb2 training set and include utterances collected from 157 male and 112 female speakers who together cover 10 different English accents. The number of trials for each accent varies between 204k and 871k. The numbers of trials for female and male speakers was 12,882 and 24,806, respectively.

\vspace{-2mm}
\subsection{Results}
\vspace{-1mm}
\textit{Do SynVox2 datasets protect speaker identity information?}
Table \ref{tab:unlink} shows EERs for the unlinkability evaluation.
The EER is 2.07\% when both enrollment and test utterances come from the authentic VoxCeleb2 database. However, when the test utterances comes from any one of the six anonymized SynVox2 databases, EERs increase to levels in the order of 30\%.
ASV score distributions in Figure~\ref{fig:unlink} confirm the effect of anonymization. The distributions for positive (same speaker) and negative (different speaker) trials show greater overlap for the ``Authentic vs. SynVox2-OHNN-bkg-utt'' anonymized setting than for the ``Authentic vs. Authentic'' baseline.
These results confirm improvements\footnote{Perfect anonymization corresponds to EERs of 50\%, i.e. fully overlapping score distributions.} to privacy through anonymization, indicating greater difficulty to link authentic and anonymized utterances/voices.

\textit{Can SynVox2 datasets be used to train an ASV model?}
The results of utility assessments are shown in Table 2.
The ASV model trained with authentic data gives an EER of 1.33\%.  Models  trained with anonymised data produce EERs of 7\% and above, with results for utterance-level embeddings being consistently superior to those for speaker-level embeddings.
These observations highlight the importance of protecting intra-speaker variation.
The introduction of additive noise through both \emph{aug} and \emph{bkg} approaches is beneficial, though EERs remain substantially higher than that of the baseline


\textit{Are the ASV models trained using SynVox2 datasets fair in terms of gender and accent? }
Figure \ref{fig:combined_fair} shows EERs and FDRs for each gender and accent groups.
The trends are similar for SynVox2 and VoxCeleb2 databases.
For instance, EERs for speakers with Spanish, Indian, and Australian accent are the highest. For gender groups, the EER for female speakers is consistently higher than those for male speakers. 
We nonetheless acknowledge that both the utility (EERs) and fairness (FDRs) degrade with the use of synthetic data.

\textit{Inter-speaker variation:}
To shed light upon the impact of inter-speaker variation, we selected 100 speakers at random and plotted their corresponding embeddings using t-SNE plots \cite{van2008visualizing} when extraction is performed using utterances from the Authentic or SynVox2-OHNN-bkg-utt databases.
The results shown in Figure \ref{fig:oo_aa} show a reduction in inter-speaker variation for the SynVox2 database. This is a likely cause of the degradation to speaker verification performance

\begin{table}[t!]
\caption{ EERs (\%) on official VoxCeleb1 test set achieved by ASV downstream models (ECAPA-TDNN) trained with different datasets.
Results in first row are reported using the ASV model trained on authetic VoxCeleb2 dataset to give an indication.
Remaining rows were obtained using different SynVox2 datasets.}
\label{tab:vox1test}
\centering
\begin{tabular}{ll}
\toprule
Training dataset     &  EER(\%) \\ \midrule 
Authentic                  & \phantom{0}1.33 \\ \midrule 
SynVox2-OHNN-spk          & 11.40\\ 
SynVox2-OHNN-aug-spk      & 10.67 \\ 
SynVox2-OHNN-bkg-spk    & 10.64\\ \midrule 
SynVox2-OHNN-utt         & \phantom{0}7.74 \\ 
SynVox2-OHNN-aug-utt     & \phantom{0}7.38  \\ 
SynVox2-OHNN-bkg-utt   & \phantom{0}7.58 \\ 
\bottomrule
\end{tabular}
 \vspace{-0.4cm}
\end{table}

\section{Conclusions}
We present in this paper our attempt to create privacy-friendly synthetic VoxCeleb2 datasets for ASV training.
By employing the OHNN-based speaker anonymization technique, we generate new SynVox2 substitute. The goal is to provide a better balance between privacy protection and utility.
We also introduce metrics for evaluation in terms of privacy, utility, and fairness. 
Results show that anonymization is reasonably successful in protecting speaker identities.
However, while the use of utterance-level embeddings and the addition of additive noise is somewhat successful in compensating for reductions to intra-/inter-speaker variation, the EER of a downstream speaker verification system increases from 1.33\% to 7\%. While this result might seem disappointing, the use of anonymized datasets may be compulsory in some settings and scenarios; increasing privacy legislation might mean that, one day, there is no alternative.  We hence expect research in this direction to continue and to attract greater attention in the future.  Further work should investigate the reduction in inter-speaker variation stemming from anonymization. Our results show that there is potential for compensation  strategies to reduce the gap in utility between authentic and privacy-friendly databases.

\noindent
\textbf{Acknowledgement}
This study is supported by JST CREST Grants (JPMJCR18A6 and JPMJCR20D3), MEXT KAKENHI Grants (21K17775, 21H04906, 21K11951, 22K21319), and the VoicePersonae project (ANR-18-JSTS-0001). Thanks Xingwei Sun for providing DeepFilter-inspired model.
\vfill\pagebreak

\bibliographystyle{IEEEbib}
\bibliography{strings,refs}

\end{document}